\documentclass{optica-article}

\journal{opticajournal} 

\articletype{Research Article}

\usepackage{lineno}

\begin{document}

\title{Degradation of $\text{Ta}_2\text{O}_5$ / $\text{SiO}_2$ Dielectric Cavity Mirrors in Ultra-High Vacuum}

\author{Alyssa Rudelis\authormark{1,*}, Beili Hu\authormark{1}, Josiah Sinclair\authormark{1}, Edita Bytyqi\authormark{1}, Alan Schwartzman\authormark{2}, Roberto Brenes\authormark{3}, Tamar Kadosh Zhitomirsky\authormark{3}, Monika Schleier-Smith\authormark{4}, and Vladan Vuleti\'c\authormark{1}}

\address{\authormark{1}Department of Physics, MIT-Harvard Center for Ultracold Atoms and Research Laboratory of Electronics, Massachusetts Institute of Technology, Cambridge, MA 02139, USA\\}
\address{\authormark{2}Department of Materials Science and Engineering, Massachusetts Institute of Technology, Cambridge, MA 02139, USA\\}
\address{\authormark{3}Electrical Engineering and Computer Science Department, Massachusetts Institute of Technology, Cambridge, MA 02139, USA.\\}
\address{\authormark{4}Department of Physics, Stanford University, Stanford, CA, 94305, USA\\}

\email{\authormark{*}rudelis@mit.edu} 


\begin{abstract*} 
In order for optical cavities to enable strong light-matter interactions for quantum metrology, networking, and scalability in quantum computing systems, their mirrors must have minimal losses. However, high-finesse dielectric cavity mirrors can degrade in ultra-high vacuum (UHV), increasing the challenges of upgrading to cavity-coupled quantum systems. We observe the optical degradation of high-finesse dielectric optical cavity mirrors after high-temperature UHV bake in the form of a substantial increase in surface roughness. We provide an explanation of the degradation through atomic force microscopy (AFM), X-ray fluorescence (XRF), selective wet etching, and optical measurements. We find the degradation is explained by oxygen reduction in Ta$_2$O$_5$ followed by growth of tantalum sub-oxide defects with height to width aspect ratios near ten. We discuss the dependence of mirror loss on surface roughness and finally give recommendations to avoid degradation to allow for quick adoption of cavity-coupled systems.

\end{abstract*}

\section{Introduction}

High-reflectivity, low-loss optical mirrors are the foundation for high-finesse optical cavities, which are indispensable tools for studying fundamental light-matter interactions and quantum information science \cite{Park2022,Welte2018,Hijlkema2007,Ramette2022, Haas2014}. Cavity QED systems can mediate entanglement generation in many-particle systems for both metrological and quantum simulation applications \cite{Hartmann2006,Pezze2018}. Indeed, cavity-mediated particle-particle interactions allow for large ensembles to be entangled, enabling measurement resolutions beyond the standard quantum limit (SQL) \cite{Greve2022,Hosten2016,Pedrozo2020}. Further, optical cavities can act as nodes for coherent photon transfer in quantum networks of various qubit hardware platforms, from NV-centers, to trapped ions, to neutral atoms \cite{Ritter2012, Brekenfeld2020, Christoforou2020,Schroder2017}. Cavities can even aid the path to fault-tolerant quantum computing: at some point, the quantum computing architecture of neutral atom arrays will reach a critical point in system size and atomic qubit state readout speed. By connecting smaller modules and by mediating nondestructive microsecond-scale state readout through cavity measurements, cavities can address both of these challenges \cite{Ramette2022, Teper2006,Deist2022}. To achieve these proposed benefits across metrology, quantum simulation, and quantum information science, a cavity-qubit system must be well within the strong-coupling regime. To be in this regime with, for example, the neutral atom qubits involved in this work, the system must have a large single-atom cooperativity, $\eta=4g^2/\kappa\Gamma\geq1$, which depends both on atomic ($\Gamma$) and cavity ($\kappa$) losses and where g is the single-atom Rabi frequency \cite{Tanji-Suzuki2011}. Thus, for a given atomic species, qubit number, and cavity mode volume, reducing cavity mirror loss and transmission is critical to reaching the strong-coupling regime and taking full advantage of optical cavities' capabilities \cite{Kong2021}.

High-reflectivity optical cavity mirrors near the infrared wavelengths needed to interact with alkali atoms are typically implemented as dielectric stacks with alternating layers of $\text{Ta}_2\text{O}_5$ (n=2.04) and SiO$_2$ (n=1.45). To reach the required reflectivities (1-50ppm transmission with $\sim$1ppm loss), these stacks consist of twenty to fifty layers of these materials, with each layer's thickness determined by the design optical wavelength(s) \cite{Gao2020}. These stacks are typically capped with the higher index of refraction material, Ta$_2$O$_5$, in order to reach the largest index differential between the dielectric stack and the environment, and therefore the highest reflectivities. 

To build an atomic system within an optical cavity, the mirrors need to be UHV-compatible, often down to pressures as low as 10$^{-10}$ Torr. In order to reach such a low pressure, the entire system must be annealed at temperatures approaching 200$^\circ$C. Unfortunately, the optical properties (thickness and refractive index) of the standard mirror top layer $\text{Ta}_2\text{O}_5$ degrade in ultra-high vacuum (UHV) conditions, supposedly because the UHV pressure drops below the vapor pressure of oxygen in the material, thus resulting in oxygen reduction \cite{Gangloff2015, Cetina2013, Sterk2012, Brandstätter2013}. Previous studies show that this vacuum-induced oxygen loss can be reversed by immersion in a high-O$_2$ environment and prevented by depositing a thin (1-2 nm) cap-layer of SiO$_2$ in most scenarios. However, if a mirror is heated above 150$^\circ$ while in vacuum, the degradation is irreversible. The material mechanism that relates oxygen reduction in Ta$_2$O$_5$ to optical loss and the reason for irreversible loss after a higher temperature anneal are not well understood.

In this work, we observe an increase in cavity mirror losses after a twelve-day UHV bake at 180$^\circ$C. To identify the cause of the degradation, we utilized several diagnostic tools: optical metrology (microscopy and profilometry), atomic force microscopy, and X-ray fluorescence (XRF) spectroscopy. Additionally, we attempted several recovery steps to reverse the loss: mirror re-alignment, mirror cleaning, and selective wet-etching. Mirror re-alignment and cleaning does not recover the initial finesse of our cavity, and optical metrology does not yield an explanation for the increase in loss through either surface scratches or contaminants. AFM images of the mirror surfaces reveal a large increase in surface roughness of high-reflectivity cavity mirrors after UHV bake. To identify the chemical nature of the roughness increase, we use XRF to find the elemental composition and use selective wet-etching to detect the presence of SiO$_2$. We conclude that the degradation is caused by oxygen reduction in the top Ta$_2$O$_5$ layer followed by nucleation and growth of tantalum sub-oxide defects. Mirror losses depend on surface roughness and the optical loss scales as expected. We recommend inspecting high-finesse mirrors via AFM prior to installation and capping them in SiO$_2$ and  to avoid vacuum-induced losses. 

    \begin{figure}
        \includegraphics[scale=0.4]{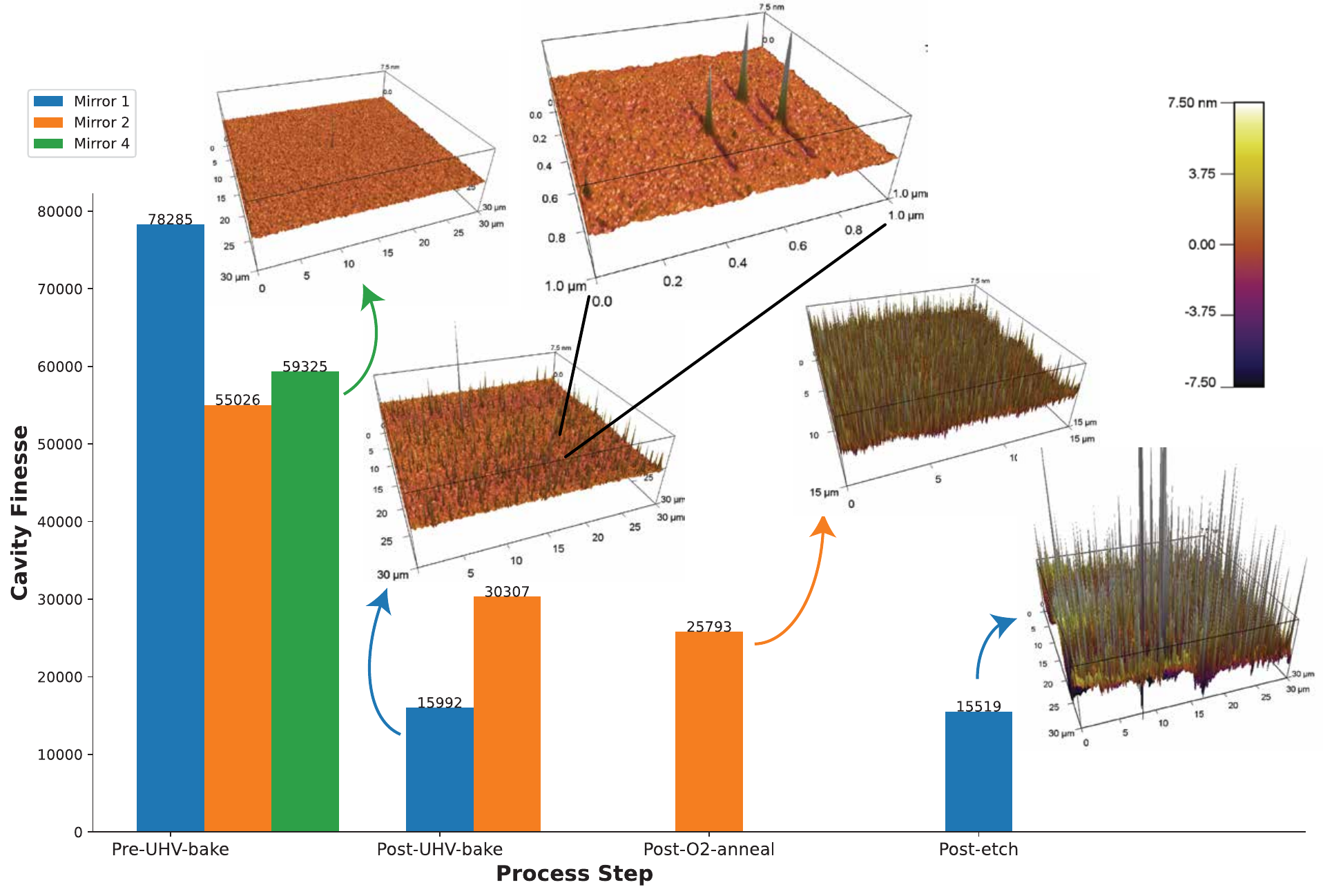}
        \caption{\label{fig:finessevprocess} Cavity finesse as a function of mirror processing step. The cavity finesses decreased after the initial vacuum anneal and after each subsequent recovery attempt that included an anneal step. Also shown are corresponding AFM images, all scaled equivalently, of the top surface of the mirror coatings.} 
        \end{figure}

\section{Materials and Methods}

    \subsection{Mirror Coating Runs}
    
    The mirrors involved in this study come from two distinct coating runs--both from Advanced Thin Films (ATF) in Boulder, CO in 2015. The differences between these coatings are as follows: dielectric stack thicknesses, coating runs, and defect presence before vacuum installation. Both dielectric stack coatings were designed for 780nm and 1560nm wavelength light. Coating Run I was designed for transmission of 2ppm at 780nm and has a total thickness of 11.6$\mu$m. Coating Run II was designed for transmission of 50ppm at 780nm with a total thickness of 7.8$\mu$m. To meet these specifications, the two mirror coatings have a different number of layers in the dielectric stack, but the substrates (UV Fused Silica), layer types (Ta$_2$O$_5$ / SiO$_2$), and layer orders (capped in Ta$_2$O$_5$) are otherwise identical. Both coating runs were completed using ion beam sputtering (IBS) followed by a high-temperature anneal. Under atomic force microscopy (AFM) inspection, we discovered at least one of the non-baked mirrors from Coating Run I had point defects with a density, $\rho_0\sim11\times10^3$mm$^{-2}$ (density of point defects after anneal, $\rho_f\sim237\times10^3$mm$^{-2}$). We found no point defects on non-baked mirrors from Coating Run II. Additionally, mirrors with Coating Run II did not degrade after the initial vacuum anneal, so this study is focused on documenting and diagnosing a failure mechanism for optical degradation of mirrors with Coating Run I. As such, all quoted finesses in this work are calculated from optical ringdown measurements of a cavity comprised of the mirror-under-test (always from Coating Run I) and a non-degraded mirror from Coating Run II as the output coupling mirror.
    
    \subsection{Annealing and Optical Methods}
    
    The initial UHV bake of the dielectric mirrors in this study was conducted in a stainless steel and fused silica vacuum chamber used for experiments with rubidium atoms. We installed two, two-mirror cavities in the chamber through the upper view port of the system. Each cavity had different properties, with one symmetric cavity comprised of two mirrors from Coating Run I ("Mirror 1" and "Mirror 3") and one asymmetric cavity comprised of one mirror from Coating Run I ("Mirror 2") and one mirror from Coating Run II. Both cavities had identical geometric properties with the length, L = 4.39cm and the radius of curvature of the mirrors, ROC = 2.5cm. Once we installed the cavities in the chamber, we sealed all conflat flanges with the recommended torque and pumped the system out following a standard pumping procedure including turbo, ion, and titanium sublimation pumping stages. During this process, we heated the chamber to a temperature of 180$^\circ$C using fiberglass heating tapes wrapped around the metal surfaces. We held the chamber at this temperature for twelve days. The pressure at the beginning of this period was $2\times 10^{-3}$ Torr and at the end of the period was $3\times 10^{-10}$ Torr as indicated by the ion pump current.

    To perform ringdown time measurements of the cavities throughout this process, we had two distinct cavity setups: 1) ringdown measurements of the symmetric and asymmetric cavities described above in vacuum over the course of the bake and 2) ringdown measurements of each mirror from coating run I with a mirror from Coating Run II as the output coupling mirror for post-mortem measurements. For all finesse measurements quoted for a single mirror in Fig.~\ref{fig:finessevprocess}, the finesse is calculated for a cavity comprised of the mirror-under-test and a mirror from Coating Run II as the output coupling mirror. 
    
    In both cavity measurement setups, we focused resonant 780nm light onto the cavity mode using mode matching lenses and steering mirrors and triggered an acousto-optic modulator (AOM) with a high optical signal at the output of the cavity. This trigger causes the AOM to switch off the cavity input light as a cavity piezoelectric transducer (PZT) scans the length of the cavity across laser resonance. To determine the characteristic 1/e ringdown time, $\tau$, we fit the resulting decay of optical power at the cavity transmission photodiode with an exponential function and extract the time constant. With the ringdown time and the cavity free-spectral range, defined as $\text{FSR}=\frac{c}{2L}$, where L is the distance between cavity mirrors, we calculate cavity finesse using $\mathcal{F}=\text{FSR}/\kappa$ where $\kappa=1/(2\pi\tau)$ is the cavity linewidth in Hz \cite{Tanji-Suzuki2011}. 
    
    Prior to the UHV bake but after installation in the chamber, the ringdown times of the symmetric and asymmetric cavities were 4.64\textmu s and 2.56\textmu s, corresponding to finesses of 9.9$\times$10$^4$ and 5.5$\times$10$^4$, respectively. The finesses for individual mirrors are plotted in Fig.\ref{fig:finessevprocess} and again are calculated assuming the asymmetric cavity arrangement comprised of one mirror from Coating Run I and one mirror from Coating Run II. After heating the mirrors during our UHV bake, initial ringdown time measurements in the chamber were less than the symmetric and asymmetric cavity transmission photodetectors' response times and thus could not be quantified using our exponential fit method. Additionally, the modes of both cavities were visibly asymmetric on transmission cameras after the UHV bake, indicating astigmatism and mirror misalignment. Misalignment indicates poor matching to the TEM00 mode of an optical cavity and reduces the overall efficiency of optical coupling into the cavity mode. This reduction in coupling efficiency effectively increases the losses of the cavity system. After removing the cavities from the chamber, re-aligning, and cleaning the mirrors, the finesses via ringdown time measurements were still significantly less than pre-UHV-bake, as depicted in the "post-UHV-bake" bar in Fig.~\ref{fig:finessevprocess}.
    
    Specifically because we did not analyze the surfaces and materials of the mirrors before baking, for a controlled comparison, we include the finesse of a cavity comprised of Mirror 4, a non-baked mirror from Coating Run I, and an output coupling mirror from Coating Run II. This Mirror 4 cavity mirror is critical for surface roughness comparisons before and after the bake. We exclude finesse measurements of Mirror 3 because we did not characterize a well-aligned cavity of this mirror in a pair with a mirror from Coating Run II before the UHV bake. Further, Mirror 3 was irreparably scratched during initial cleaning recovery efforts after the UHV bake.
    
    Following minimal improvements to cavity finesse via mirror cleaning and finding no culpable scratches or contaminants on the mirror surfaces via microscopy and profilometry, we adopted the mirror recovery procedure from \cite{Gangloff2015}. In this previous study of mirror degradation in vacuum conditions, an O$_2$ anneal successfully recovered mirrors by removing oxygen vacancies that developed during a UHV bake. This method was successful in mirrors UHV-baked at temperatures below 150$^\circ$C, but it did not recover mirrors that were UHV-baked at higher temperatures \cite{Gangloff2015}. Our attempted recovery anneal was carried out in a 100\% O$_2$ tube furnace at 400$^\circ$C for four hours, only on Mirror 2. The resulting finesse measurement indicates further degradation of Mirror 2's reflective surface after the O$_2$ anneal, as shown in Fig.~\ref{fig:finessevprocess}. Discovering persistent degradation of our mirrors despite cleaning and O$_2$ annealing led us to pursue systematic, materials characterization of each mirror from Coating Run I beyond the scope of previous studies. The summary of the material recovery efforts in terms of finesse are displayed in Fig.~\ref{fig:finessevprocess} and the material characterization results are summarized in Figs.~\ref{fig:peakheightdistributions}-\ref{fig:XRF}. Descriptions of the materials-focused work follows in Sec.~\ref{nonoptical}.  
   
    \subsection{Non-Optical Methods}
    \label{nonoptical}

    We utilized three main non-optical characterization methods to diagnose and understand the loss mechanism in our mirrors: atomic force microscopy (AFM), X-ray fluorescence (XRF), and selective etching. After several rounds of cleaning and re-aligning, the cavity mirrors did not recover their initial, pre-UHV-bake performance. Thus, we postulated that physical changes to the surface of the mirror could be the source of loss. To verify, we measured mirror roughness using AFM, a method that maps the topology of a surface using deflections of and active feedback to a microscopic cantilever. In our case, we used an Asylum Research Cypher AFM machine in repulsive tapping-mode with an aluminum-coated cantilever. After mapping all of our mirrors' HR surfaces and noting a roughness increase after both the UHV bake and O$_2$ anneal, we moved on to identify the defects. To do so, we carried out two material analyses: XRF and selective wet etching. We used a Bruker Tracer-III handheld XRF device to qualitatively determine which elements were present in the reflective coating of our mirrors, both before and after UHV bake. The XRF results showed that the increase in roughness was not due to contamination of the mirror dielectric stack with an external material. We then further narrowed down the identity of the growths by using a selective wet etch of the surface. In the semiconductor industry, a common chemical etch, or method to remove material from a surface in a controlled manner, is a buffered oxide etch (BOE). BOE etches SiO$_2$ at a rate orders of magnitude faster than pure silicon or Ta$_2$O$_5$. For this reason, we used selective wet-etching with a 7:1 buffered oxide etch (BOE) in the MIT.nano cleanroom to determine whether the defects were SiO$_2$ or another material. The BOE was unable to remove the defects.

\section{Results and Discussion}

    \subsection{Process Effects on Cavity Finesse}         

      \begin{figure}
          \centering
          \includegraphics[scale=0.8]{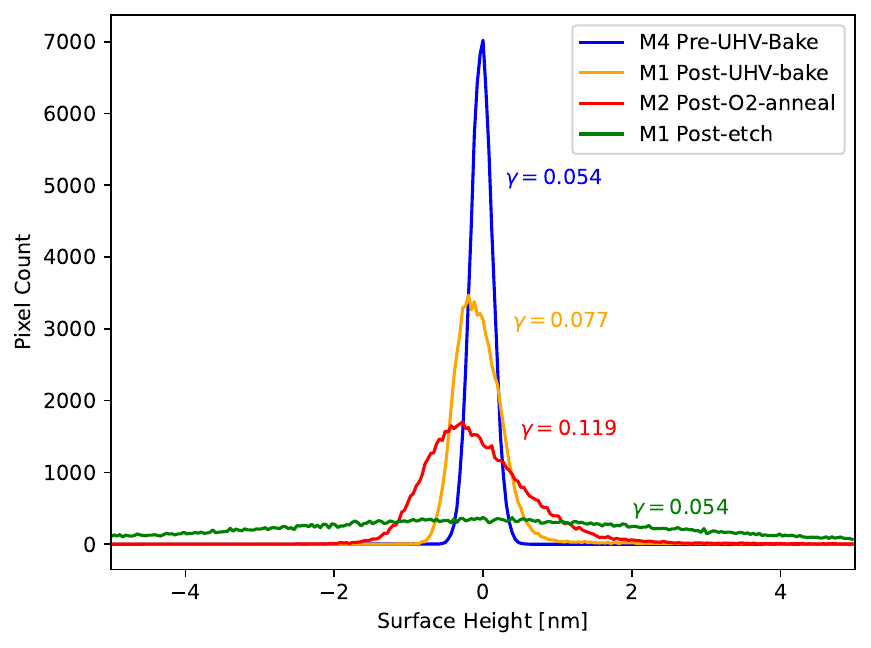}
          \caption{Histogram of mirror surface heights after the various processing steps. The distribution in surface height of a single mirror becomes less Gaussian with each heating step, indicated by increasing skew, $\gamma$. Additionally, each processing step intended to return the mirror surface roughness to the pre-UHV-bake value instead increased the deviation in mirror roughness.}
          \label{fig:peakheightdistributions}
      \end{figure}

    Process effects on our cavity finesse were the first indicator of material degradation that led us to pursue more thorough mirror characterization methods. As discussed previously, the drop in optical quality was only observed in mirrors with Coating Run Run I, so our analysis focuses on those mirrors. The initial UHV bake caused a severe decline in cavity finesse, as shown in the comparison of the first two data points in Fig.~\ref{fig:finessevprocess}. The cavities were designed to be in the strong coupling regime with a single atom cavity cooperativity $\eta\geq1$ for the rubidium D$_2$ line. Before the UHV bake, $\eta=3.4$ and with the drop in finesse of mirrors from Coating Run Run I, $\eta=0.9$ and was no longer in the strong-coupling regime. In an attempt to recover the initial finesse of the cavities, we followed the recommendation of a previous study that claimed oxygen vacancy formation in the  Ta$_2$O$_5$ cap \cite{Gangloff2015}. We annealed Mirror 2 from Coating Run Run I in a pure oxygen environment at 400$^\circ$C for four hours. This did not recover the initial finesse, implying in agreement with the previous study that an additional mechanism is responsible for vacuum mirror degradation at temperatures above 150$^\circ$C. Our subsequent recovery attempt using a BOE on Mirror 1 was also unable to improve the degraded finesse.
    
    \subsection{Process Effects on Mirror Roughness}  
      
        \begin{figure}
        \includegraphics[scale=0.75]{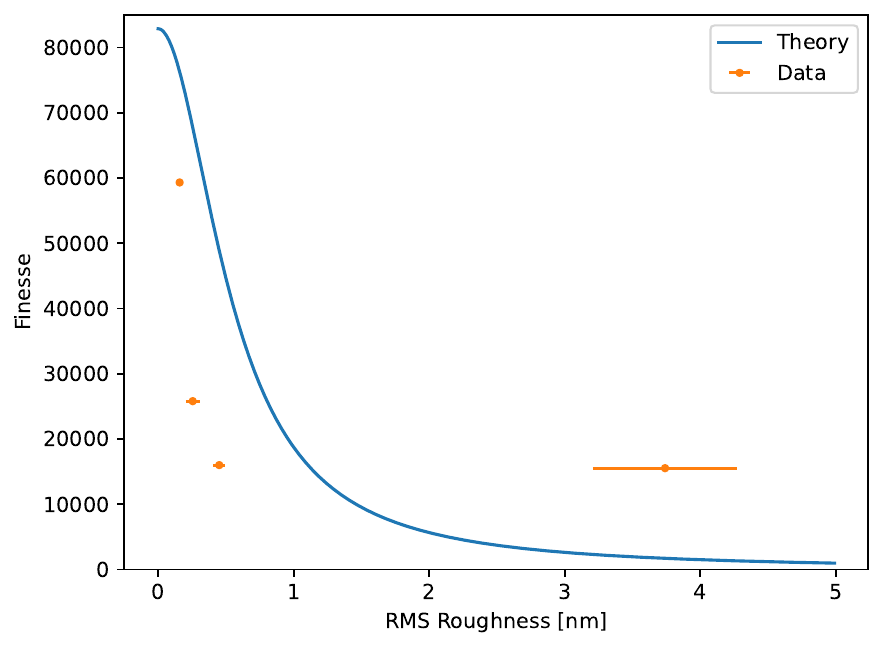}
        \caption{\label{fig:rough_v_finesse} Expected theoretical reduction in cavity finesse with mirror surface roughness compared with measured reduction in cavity finesse with mirror surface roughness. The theory curve is calculated with no free parameters.}
        \end{figure}

    AFM images of the dielectric surface of UHV-baked cavity mirrors from Coating Run I show a drastic increase in surface roughness as compared with images of non-baked mirrors, as shown in Fig.~\ref{fig:finessevprocess}. The roughness is attributable to an abundance of high-aspect-ratio point defects after the anneal. Non-baked mirrors from Coating Run I do exhibit these point defects, but at a much lower density than their baked counterparts. Mirrors from Coating Run II never display these point defects, before or after the UHV bake. This discrepancy between the behavior of the two coating runs implies that manufacturing errors in Coating Run I could be partially responsible for the severe degradation under UHV bake. In addition to the deviation in mirror defect heights increasing with each processing step, as shown in Fig.~\ref{fig:peakheightdistributions}, the distribution of defect heights also becomes more skewed and less Gaussian after each heating processing step. The post-BOE curve does not follow this increase in skewness because the distribution of defect heights is so uniform that the error dominates. The root-mean-square (rms) surface roughness extracted from the AFM measurements of mirrors from Coating Run I plotted in Fig.~\ref{fig:rough_v_finesse} shows that the increase in roughness with each processing step tracks the decrease in cavity finesse during the same period. Indeed, one would expect the loss of the mirrors to increase with surface roughness as \cite{Bennett1961}
    
    \begin{equation}
        L=L_0+\frac{(4\pi\sigma_{\text{rms}})^2}{\lambda^2}.
    \end{equation}

    We can relate the losses of individual mirrors to the finesse of an optical cavity with $\mathcal{F}=2\pi/(L_1+L_2+T_1+T_2)$ \cite{Li2022}. Combining these equations describes the relationship between our cavity finesse and mirror roughness as
    
    \begin{equation}
        \mathcal{F}=\frac{2\pi}{L_1+L_2+\text{exp}[-(4\pi\sigma_{\text{rms}})^2/\lambda^2]+T_1+T_2},
    \end{equation}
    
    where $L_1=10$ppm and $L_2=10$ppm are the initial losses, $T_1=50$ppm and $T_2=5.8$ppm are the measured transmission coefficients of cavity mirrors from Coating Run II and I, respectively, and we have an added loss term that depends on $\sigma_{\text{rms}}$ \cite{Straniero2015}. Our finesse and roughness data from the optical and AFM measurements are plotted in Fig.~\ref{fig:rough_v_finesse} along with this theoretical curve. The horizontal error bars indicate one standard deviation of error in our measurements of roughness, calculated from a small number (1-4) of AFM scans. The number of scans was limited by time, cantilever tips, and the curvature of the HR mirror surfaces. The vertical error bars for finesse are smaller than the size of the data points. We find good qualitative agreement between data and theory and conclude our data are consistent with mirror roughness as the dominant loss mechanism in the degraded mirrors. The final data point at highest roughness in this plot strays from the theory the most, lying far outside one standard deviation of error in roughness and finesse. We would not expect perfect quantitative agreement between our measurements and the model due to the skewness of the defect height distributions. Further, our inability to perform AFM and ring-down measurements on the exact same point on the mirror surface due to a lack of location-identifying markers on the mirror surface contributes to systematic error. Finally, the redistribution of defect sizes between processing steps could indicate a higher level of periodicity in the surface topology, potentially introducing interference phenomena and complicating the simple model of cavity finesse dependence on surface roughness plotted in Fig.~\ref{fig:rough_v_finesse}. 

        \begin{figure}
        \centering
            \includegraphics[scale=0.75]{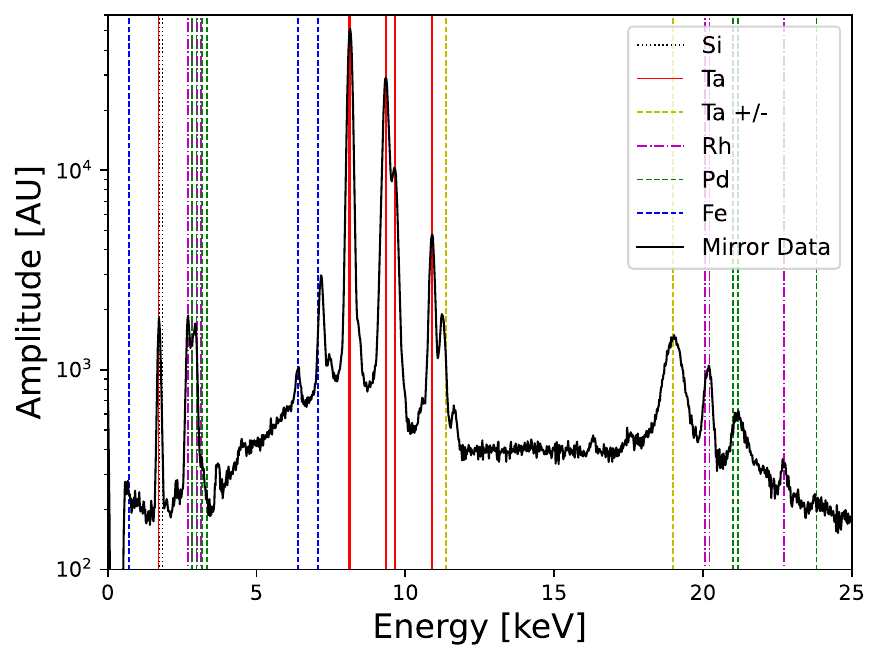}
            \caption{X-ray fluorescence rules out the possibility that a contaminant from the vacuum chamber caused the mirror surface roughness increase. Only Si, Ta, and artifacts of the XRF device (Rh, Pd, and Fe) were resolvable in the measurement.}
        \label{fig:XRF} 
        \end{figure}

    \subsection{Elemental Analysis}
    To further understand the degradation mechanism, we set out to identify the chemical composition of the defect growths.
    
        \subsubsection{X-Ray Fluorescence Spectroscopy}
        To determine the identity of individual elements in the surface point defects in the  AFM images in Fig.~\ref{fig:finessevprocess}, we analyzed each mirror surface with XRF, yielding an identical set of characteristic peaks as shown in Fig.~\ref{fig:XRF}. The most prominent fluorescence peaks were those characteristic of tantalum, followed by silicon, palladium, rhodium, and iron, the last three of which are elements found in the XRF detector itself. This measurement confirms that the defects must originate from the materials within the dielectric stack instead of from a contaminant. Thus, the XRF implies the defects are pure silicon, pure tantalum, or oxides of those elements. 
        
        \subsubsection{Selective Wet-Etching}
        A selective BOE further narrows down the identity of the defect: BOE is highly selective for SiO$_2$ over the other material possibilities. Specifically, BOE etches SiO$_2$ at a rate of 2nm/s, Si at a rate of 0nm/s, and Ta$_2$O$_5$ at a rate of 3$\times$10$^{-4}$ nm/s at room temperature and pressure \cite{Braverman2018, Christensen1999}. Thus, etching our HR mirror surface with BOE reveals whether the roughness increase was due to SiO$_2$ or one of the other dielectric materials in our mirrors. At room temperature, a BOE of fourteen seconds should etch away at least 16nm of SiO$_2$, which is not what we observe \cite{Proksche1992}. Because the height of the defects did not decrease after the BOE, we conclude the defects are not SiO$_2$. However, as is shown in the post-BOE AFM in Fig.~\ref{fig:finessevprocess}, the roughness of Mirror I did not remain constant, but increased after the BOE step. We explain this unintuitive result by noting that additional heating is required prior to the BOE in order to apply a photomask to protect the AR mirror coating. This additional annealing likely caused further defect growth. Given that the roughness of each mirror increased during any processing step that elevated temperatures, the increase of roughness after etching is consistent with our other observations. It is also conceivable that the defect growths from the UHV bake produced holes in the top Ta$_2$O$_5$ layer (initial thickness of 126nm), exposing the subsequent SiO$_2$ to the BOE. This, too, would help explain the apparent "growth" of Ta$_2$O$_5$ defects after BOE by eating away at their floor.

        \subsubsection{Thermodynamic Considerations}
        After narrowing down the defect identity by ruling out contaminants and SiO$_2$, the final defect determination can be made through thermodynamics--specifically with the help of an Ellingham Diagram. Ellingham diagrams plot the Gibbs free energy of formation, $\Delta G^0$, of various compounds (especially oxides) at a given temperature. The more negative $\Delta G^0$, the more thermodynamically likely a chemical makeup is under the given conditions. Plotting the Ellingham Diagram for tantalum oxides at the high temperature of 180$^\circ$C encountered in this study reveals that because $\Delta G^0$ is negative and less than other tantalum oxides at that temperature and because the partial pressure of O$_2$ in our system was much less than one atmosphere, the only possible identity of the defects is Ta$_2$O$_5$ sub-oxides \cite{ellingham}.

\section{Conclusion}

In conclusion, we find that the UHV degradation of Ta$_2$O$_5$-capped dielectric cavity mirrors is not just attributable to reversible oxygen vacancy formation in the tantalum oxide layer. In fact, we find that when baked above 150$^{\circ}$C in UHV, these mirrors can undergo an irreversible nucleation and growth process of the dielectric stack materials. This material change is directly observable in optical ringdown time characterizations of a cavity and under AFM inspection of mirror surfaces. This observation confirms and extends the result from a previous study where optical quality of cavity mirrors baked at 150$^{\circ}$C could not be recovered with oxygen flow \cite{Gangloff2015}. It is likely that oxygen defects form at low temperatures in vacuum and subsequently act as nucleation points for topologically distinct growth of Ta$_2$O$_5$ sub-oxide structures at higher temperatures. We also find that the presence of low-density defects on unbaked mirrors is correlated with degradation post-UHV-bake. Mirrors from Coating Run I were the only mirrors in our study to degrade under vacuum conditions, and these mirrors were also the only mirrors with low-density defects prior to installation. One potential explanation for this difference could be that the deposition process becomes more unpredictable as dielectric stack thickness increases. Thus, small changes in the commercial IBS and annealing processes used to create these dielectric stacks could have immense impact on the longevity and performance of high-reflectivity mirrors. Further, the formation of "whisker" defects with high aspect ratios are well-documented in other species of metal-oxide materials during oxygen reduction processes \cite{Chang1984}. Although this whisker formation process has not previously been observed in Ta$_2$O$_5$ specifically, it is generally accepted that the presence of surface defects (oxygen vacancies or external defects) in transition metal oxides often are responsible for subsequent catalytic activity, which supports our hypothesis of oxygen vacancies as defect nucleation sites \cite{Huggins2005}. To prevent such material degradation under vacuum conditions, high-reflectivity optical mirrors should always be inspected for defects via AFM measurements and should be capped in a thin protective layer of SiO$_2$ before vacuum installation.

\begin{backmatter}
\bmsection{Funding}
This project was funded by NSF, NSF CUA, NASA, and MURI through ONR

\bmsection{Acknowledgments}
We wish to acknowledge the Bulovi\'c group (ONELab) at MIT for help with initial mirror cleaning, annealing, and XRD access. We also acknowledge Kurt Broderick and Maansi Patel from MIT.nano for assistance with clean room processing.

\bmsection{Disclosures}
The authors declare no conflicts of interest.

\bmsection{Data Availability Statement}
Data underlying the results presented in this paper are not publicly available at this time but may be obtained from the authors upon reasonable request.

\end{backmatter}

\bibliography{main}

\end{document}